\documentclass[epj]{svjour}
\usepackage{graphics}
	\def\Journal#1#2#3#4{{#1} {\bf #2}, (#4) #3}
	\def\PLB{{Phys. Lett.}  B}
	\def\PRL{{Phys. Rev. Lett.}}
	\def\PRB{{Phys. Rev.} B}
	\def\ZPB{{Z. Phys.} B}
	\def\et{{\it et al.}}

\input epsf.sty
\begin{document}

\title{Spin Correlations in the Two-Dimensional Spin-5/2 Heisenberg
Antiferromagnet Rb$_2$MnF$_4$}

\author{Y. S. Lee\inst{1} \and M. Greven\inst{1}\thanks{\emph{Present
address: } Department of Applied Physics and Stanford Synchrotron Radiation
Laboratory, Stanford University, Stanford, CA 94305.}
\and B.O. Wells\inst{1} \and R. J. Birgeneau\inst{1} \and G. Shirane\inst{2}
}

\institute{Department of Physics, Massachusetts Institute of Technology,
Cambridge, MA 02139 \and Department of Physics, Brookhaven National Laboratory, Upton,
NY 11973
}

\date{Received: date / Revised version: date}


\abstract{
We report a neutron scattering study of the instantaneous spin
correlations in the two-dimensional spin $S=5/2$ square-lattice Heisenberg
antiferromagnet Rb$_2$MnF$_4$.  The measured correlation lengths are
quantitatively described, with no adjustable parameters, by
high-temperature series expansion results and by a theory based on the
quantum self-consistent harmonic approximation.  Conversely, we find
that the data, which cover the range from about 1 to 50 lattice
constants, are outside of the regime corresponding to renormalized
classical behavior of the quantum non-linear $\sigma$ model.  In
addition, we observe a crossover from Heisenberg to Ising critical
behavior near the N\'{e}el temperature; this crossover is well
described by a mean-field model with no adjustable parameters.
 % end of PACS codes
} %end of abstract
\maketitle

\section{Introduction}
\label{intro}

The physics of the two-dimensional square-lattice quantum Heisenberg
antiferromagnet (2DSLQHA) continues to receive much attention.  In
addition to the basic interest in studying the role of quantum
fluctuations in this high-symmetry low-dimensional system,
experimental and theoretical efforts have heightened with the
discovery that the undoped parent compounds of high-$T_c$ superconductors are
typically very good realizations of the spin $S=1/2$ 2DSLQHA.  In
particular, neutron scattering experiments on the insulating lamellar
copper oxides La$_2$CuO$_4$ \cite{keimer-214} and Sr$_2$CuO$_2$Cl$_2$
\cite{greven} have elucidated the spin fluctuations of
the 2DSLQHA model for the extreme quantum limit of $S=1/2$.  The
measured spin-spin correlation lengths of Sr$_2$CuO$_2$Cl$_2$ are
quantitatively described \cite{greven} by quantum Monte Carlo results
based on the square-lattice nearest-neighbor Heisenberg
model~\cite{mak-ding,beard}, and by a low-temperature theory based on
the quantum non-linear $\sigma$ model developed by Chakravarty,
Halperin, and Nelson (CHN)~\cite{chn}.  The remarkable agreement
between the experimental data for $S=1/2$ and the theoretical results
of CHN, as extended by Hasenfrantz and Niedermeyer (HN)~\cite{hn}, was
found not to hold for higher spin values, $S>1/2$.  In particular, in
the $S=1$ 2DSLQHA systems K$_2$NiF$_4$~\cite{greven} and
La$_2$NiO$_4$~\cite{nakajima}, the measured correlation lengths were
found to deviate significantly from the CHN-HN prediction.  A
systematic high-temperature series expansion
study~\cite{elstner-series} showed that the deviations increase
progressively with increasing spin values above $S=1/2$.  Recent Monte
Carlo work~\cite{beard} suggests that the explanation for this
discrepancy may simply be that the series expansion and neutron scattering
results are not in the asymptotic low-temperature regime for which the
CHN-HN prediction is expected to hold.

For the 2DSLQHA, quantum spin fluctuations strongly renormalize the
spin-stiffness and spin-wave velocity in the $S=1/2$ system
Sr$_2$CuO$_2$Cl$_2$, but there appears to be no fundamental change
from classical behavior.  For $S=5/2$, the 2DSLQHA should be even less
affected by quantum renormalization and thus should correspond more
closely to a classical spin system.  Indeed, high-temperature series
expansion studies reveal a near-agreement between the correlation
lengths of the $S=5/2$ 2DSLQHA and the $S=5/2$ ferromagnetic system,
for which the ground state is classical~\cite{elstner-long}.  In
addition, upon scaling the temperature by $J_{nn}S(S+1)$, series
expansion results~\cite{elstner-series} indicate that the correlation
length of the 2DSLQHA rapidly approaches that of the classical
$S=\infty$ case as the spin value is increased progressively from
$S=1/2$ to $S=5/2$.  In order to examine further the role of quantum
fluctuations in the models, and their dependence on the spin quantum
number, one requires more experimental data for the correlation length
of systems well described by the 2DSLQHA Hamiltonian with higher spin.

\begin{sloppypar}
We report an energy-integrating neutron scattering study of the 2D
instantaneous spin-spin correlations in the $S=5/2$ material
Rb$_2$MnF$_4$.  This paper is organized as follows: Sect.~\ref{sec:2}
contains preliminary details about the Rb$_2$MnF$_4$ system and about
our measurements; Sect.~\ref{sec:3} contains our experimental results
and data analysis; in Sect.~\ref{sec:4}, we present a comparison with
various theories; and Sect.~\ref{sec:5} summarizes our results.
\end{sloppypar}

\section{Preliminary Details}
\label{sec:2}

We study a high-quality single crystal of Rb$_2$MnF$_4$, which is,
in fact, the same as that used in previous neutron
studies~\cite{birgeneau-70,cowley-magnet}.  Rb$_2$MnF$_4$ has the
K$_2$NiF$_4$ crystal structure, space group $I4/mmm$, with square
planes of MnF$_2$ separated by two intervening sheets of non-magnetic
ions.  The magnetic Mn$^{2+}$ ions ($S=5/2$) form a square lattice and
are antiferromagnetically coupled to their nearest neighbors via
super-exchange through the intervening F$^-$ ions.  The
low-temperature in-plane lattice constants are $a=b= 4.20~{\rm \AA}$,
and the out-of-plane lattice constant is $c= 13.77~{\rm \AA}$.  Over
the temperature range of the experiment, $10~{\rm K}\leq T \leq
110~{\rm K}$, the in-plane lattice constant changes by only
$\sim$0.1\%, so any concomitant temperature variation of the in-plane
exchange coupling will have a negligible effect on the spin
correlations.  The exchange coupling between nearest neighbor spins in
adjacent planes is frustrated for this body-centered tetragonal spin
structure.  As a result, any effective inter-planar coupling is at
least six orders of magnitude weaker than the coupling within the
planes, leading to the quasi-two-dimensional nature of the spin
system.  From previous neutron scattering experiments on this and
isomorphous materials, it is known that the critical scattering
consists of purely two-dimensional spin fluctuations, and no spin-wave
dispersion is observable along the ${\bf c}$-direction, confirming
that a 2D spin model is
appropriate~\cite{birgeneau-k2nif4,birgeneau-quasielastic,birgeneau-77}.

The physics of the Mn$^{2+}$ square lattice is well described by
the simple spin Hamiltonian
\begin{equation}
H =  J_{nn} \sum_{<i,j>} {\bf S}_i \cdot {\bf S}_j ~+~  \sum_i g_i
\mu_B H_i^A S_i^z,
\end{equation}
where the staggered anisotropy field $H_i^A$ represents the effect of
the dipolar anisotropy which favors spin alignment along the {\bf
c}-axis.  From both NMR measurements of the sublattice
magnetization~\cite{dewijn} and neutron scattering measurements of the
spin-wave dispersion~\cite{cowley-spinwave,hamer}, one obtains
$J_{nn}=7.36\pm0.10$~K.  The small Ising anisotropy in this
predominately Heisenberg Hamiltonian is $\alpha_I=g
\mu_B H^A / \sum_{j=nn} J_{nn}S_j\simeq0.0047$, as deduced from antiferromagnetic
resonance and inelastic neutron scattering measurements of the
low-temperature spin-wave gap at the magnetic zone
center~\cite{dewijn,cowley-spinwave}.  A small three-dimensional (3D)
coupling does exist as evidenced by the transition to 3D long-range
order which occurs at $T_N=38.4$~K.  Such a 3D coupling usually
appears as a coupling between next-nearest-neighbor ($nnn$) planes.
However, the weakness of this coupling is evident from magnetic Bragg
diffraction, which reveals two domains with different stacking
arrangements of ordered MnF$_2$ planes: one in which $nnn$ planes are
ferromagnetically aligned and another in which $nnn$ planes are
antiferromagnetically aligned~\cite{birgeneau-70}.  This 3D order
reflects primarily 2D correlations with Ising critical behavior, since
the staggered magnetization of both stacking domains was found to
follow the same power law, $M_s=(1-T/T_N)^{0.16}$~\cite{birgeneau-70};
the observed critical exponent $\beta$ for the order parameter is much
closer to the 2D Ising model value $\beta=\frac{1}{8}$ than to the
conventional 3D values of approximately $\frac{1}{3}$.

The experiments were performed on the H4M thermal neutron spectrometer
at the Brookhaven High Flux Beam reactor.  The Rb$_2$MnF$_4$ crystal
was aligned with a (1$\bar{1}$0) axis perpendicular to the scattering
plane, thus having the magnetic zone center wave vector ($\frac{1}{2},
\frac{1}{2}, 0)$ and the ${\bf c}$-axis in the scattering plane.  The
spectrometer was operated in the two-axis energy-integrating
configuration with collimator sequence
$20^{\prime}$-$20^{\prime}$-sample-$20^{\prime}$.  The scattering
geometry was chosen so that outgoing neutrons were perpendicular to
the MnF$_2$ planes, thus integrating over energy at constant in-plane
momentum transfer ${\bf Q_{\rm 2D}}$ \cite{birgeneau-quasielastic}.
In this geometry, the intensity of the detected neutrons is
proportional to the static structure factor,
\begin{equation}
S({\bf Q_{\rm 2D}}) \simeq \int_{-\infty}^{E_i}S({\bf Q_{\rm 2D}},
\omega) d\omega,
\end{equation}
where ${\bf Q_{\rm 2D}}$ is the momentum transfer within the 2D
MnF$_2$ sheets, and $\omega$ is the energy transfer.  Here, we assume
that any variation in the Mn$^{2+}$ form factor has a negligible
effect in the integration over $\omega$.  Two incident neutron
energies, $E_i=14.7$~meV and $E_i=41$~meV, were used to verify that
the experiment integrates over the relevant dynamic fluctuations
properly.  Both incident energies are more than an order of magnitude
larger than the magnetic energy scale, $J_{nn}$, in Rb$_2$MnF$_4$.  By
integrating over all energies, we obtain information about the
instantaneous (equal-time) spin correlations of the system.

\section{Experimental Results and Analysis}
\label{sec:3}

We show in Figs.~1 and 2 representative two-axis scans for
$E_i=41$~meV and $E_i=14.7$~meV, respectively.  In order to extract
the intrinsic peak widths and amplitudes of the scattering, we fit our
data to the form
\begin{eqnarray}
S(q_{\rm 2D}) & = & ~{\rm sin}^2(\phi)~~
\frac{S_\parallel(0)}{1+q_{\rm 2D}^2/\kappa^2_\parallel} \nonumber \\
              &   & +~~(1+{\rm cos}^2(\phi))~~
\frac{S_\perp(0)}{1+q_{\rm 2D}^2/\kappa^2_\perp} 
\end{eqnarray}
convolved with the instrumental resolution function (which is drawn in
the top panels of the figures) on top of a sloping background.  Here,
${\bf q}_{\rm 2D}~=~(Q_x-\frac{1}{2},~Q_y-\frac{1}{2},~0)$ represents
the displacement from the center of the rod of 2D scattering measured
in reciprocal lattice units, and $\phi$ is the angle subtended by {\bf
Q} and the {\bf c}-axis.  The correlation length
$\xi_{\parallel,\perp}$ is the inverse of the {\bf Q}-space width
$\kappa_{\parallel,\perp}$.  For $T~{\Large
\stackrel{_>}{_\sim}}~1.2~T_N$, the profiles are well described by a
single 2D Lorentzian form, that is, $\kappa_\parallel=\kappa_\perp$.
As the system is cooled to the immediate vicinity of $T_N$, the
scattering exhibits a crossover from Heisenberg to Ising
behavior~\cite{birgeneau-77}.  Accordingly, we find that within about
20\% of $T_N$, for $T>T_N$, one must describe the lineshape with two
components: one 2D Lorentzian with diverging $\xi_\parallel$ and
$S_\parallel(0)$, corresponding to the longitudinal ($\|~{\bf c}$)
Ising fluctuations, and a second 2D Lorentzian corresponding to the
non-divergent transverse ($\perp$~{\bf c}) spin fluctuations.  For our
experimental configuration, the geometrical factors sin$^2(\phi)$ and
1+cos$^2(\phi)$ are both close to unity; specifically, for
$E_i=14.7$~meV, sin$^2(\phi)\simeq0.96$ and 1+cos$^2(\phi)\simeq1.04$.
The transverse scattering contribution
$S_\perp(0)/(1+q_{2D}^2/\kappa_\perp^2)$ is denoted by the dashed line
in the bottom two panels of Fig.~2.

%%==============================================================================
\begin{figure}
\centerline{\epsfxsize=2.9in\epsfbox{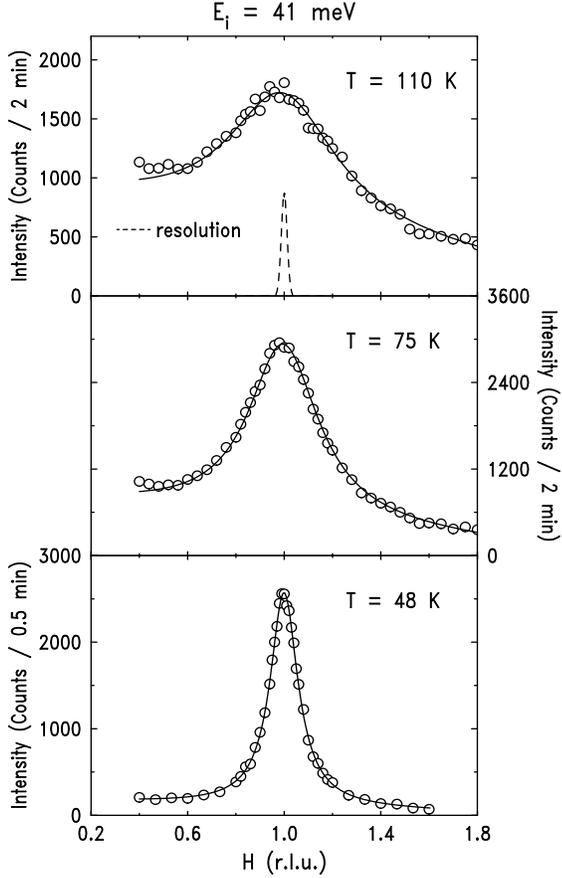}}
\vskip 5mm
\caption{
Representative energy-integrating scans along the direction
($H/2$,~$H/2$,~$L$), with $L$ chosen such that {\bf k}$_f~\|$~{\bf c},
for an incident neutron energy of 41 meV.  The solid lines result from
least-squares fits of the lineshape, Eq.(3), convolved with the
instrumental resolution.  The dashed line in the top panel indicates
the instrumental resolution. }
\label{Figure1}
\end{figure}
%%%==============================================================================

%%==============================================================================
\begin{figure}
\centerline{\epsfxsize=2.9in\epsfbox{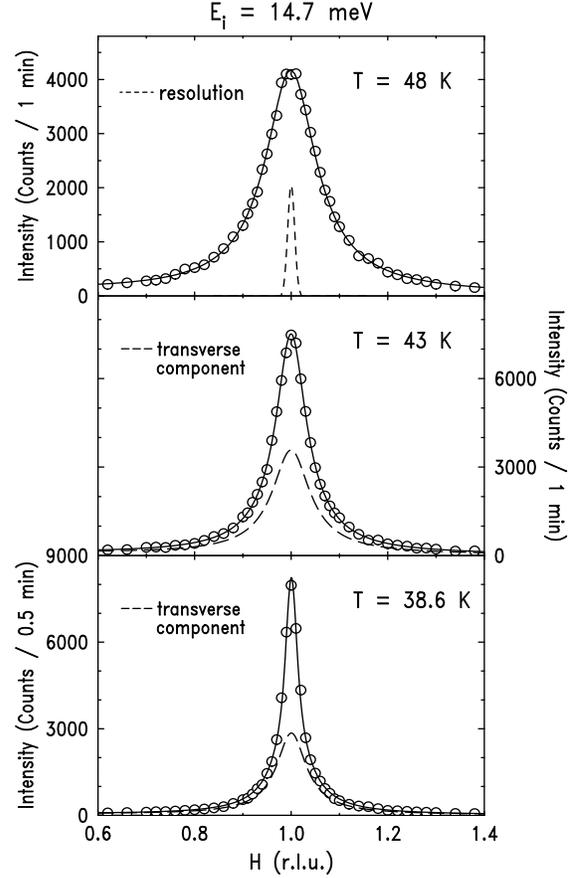}}
\vskip 5mm
\caption{
Representative energy-integrating scans along the direction
($H/2$,~$H/2$,~$L$), with $L$ chosen such that {\bf k}$_f~\|$~{\bf c},
for an incident neutron energy of 14.7 meV.  The solid lines result
from least-squares fits of the lineshape, Eq.(3), convolved with the
instrumental resolution.  The dashed line in the top panel indicates
the instrumental resolution.  The dashed lines in the bottom two
panels indicate the magnetic scattering due to the transverse spin
component.  }
\label{Figure2}
\end{figure}
%%==============================================================================

\begin{sloppypar}
In our analysis, we first fit the temperature dependence of the
transverse scattering at temperatures below $T_N$ using a very simple
model.  Within the ordered phase, the transverse spin contribution
originates from spin-wave scattering; therefore, $\kappa_{\perp}$ is
held fixed at the spin-wave value of $\kappa_\perp=0.028$ reciprocal
lattice units, while the quantity $S_\perp(0)\kappa_\perp^2$ is
assumed to increase with the Bose thermal occupation factor.  The Bose
factor, which takes into account both neutron energy gain and energy
loss due to spin-wave excitations, is $2n(\omega)+1$ where
\mbox{$n(\omega)=1/[\rm{exp}(\omega/T)-1]$} and $\omega=7.2\pm0.2$~K is the
low-temperature spin-wave gap at the magnetic zone
center~\cite{cowley-spinwave}.  Below $T_N$ we also observe a sharp
resolution-limited quasi-2D Bragg component along
$(\frac{1}{2},~\frac{1}{2},~L)$ which grows in intensity with
decreasing temperature.  This 2D Bragg scattering probably originates
from 2D sheets at the interfaces between the two different 3D stacking
domains.  The 2D Bragg scattering is assumed to have the same
temperature dependence as the 3D order parameter.  This simple model
is found to describe the data below $T_N$ very well.
\end{sloppypar}

Since the focus of our experiments is on the behavior above $T_N$, the
main impetus for measuring and modeling the scattering below $T_N$ is
to allow us to estimate the non-critical transverse scattering in the
Ising critical regime above $T_N$.  By definition, the transverse
scattering must become identical to the longitudinal scattering in the
Heisenberg regime well above $T_N$.  Therefore, in the fitting, we
assume that for the transverse component, $\kappa_\perp$ and
$S_\perp(0)\kappa_\perp^2$ can be simply interpolated linearly between
the fitted values at $T_N=38.4$~K and those at 46~K, and that above
46~K the two Lorentzians are identical, that is, within the errors the
system shows pure Heisenberg behavior.  Thus, in the final data
analysis above $T_N$, the only free fit parameters are
$\kappa_\parallel$ and $S_\parallel(0)$, which describe the Heisenberg
behavior at high-temperatures and the Ising critical scattering for
temperatures near $T_N$.  The results are shown in Figs.~3 and 4.  For
the fitted values of $\kappa_\parallel$ shown in Fig.~3, the plotted
error bars correspond to the larger of three statistical standard
deviations or one-tenth of the instrumental resolution.  Similarly,
for the $S_\parallel(0)$ data shown in Fig.~4, the plotted error bars
are equal to three standard deviations.  We note that the data
obtained with $E_i=14.7$~meV and $E_i=41$~meV agree well with each
other, thus confirming the validity of the quasi-elastic
approximation~\cite{birgeneau-quasielastic}.  The dash-dotted lines
represent the interpolated temperature dependences of the transverse
scattering parameters.  We find that varying the upper temperature
limit of the interpolation of the transverse fluctuations around 46~K
by several degrees does not change the results for $\kappa_\parallel$
and $S_\parallel$(0) within the errors.  Furthermore, we also obtain
closely similar results using a very different interpolation scheme
for $\kappa_\perp$ and $S_\perp(0)\kappa_\perp^2$ similar to that
employed in Ref.~\cite{birgeneau-77}.  The solid curve in Fig.~3
corresponds to a theory for the 2DSLQHA by Cuccoli \et~\cite{cuccoli},
which we plot with a modification incorporating the effects of the
Ising anisotropy.  We will elaborate upon this in the next Section.
The dashed line in Fig.~3 is the prediction from high-temperature
series expansion for the Heisenberg model with
$S=5/2$~\cite{elstner-series}.  In Fig.~4, the solid line is the
unmodified result from Ref.~\cite{cuccoli}, and the dashed line is the
series expansion result for the $S=5/2$ 2DSLQHA~\cite{elstner-series}.
Since the neutron scattering intensity is not measured in absolute
units, we choose a scale for the plot of $S_\parallel(0)$ in Fig.~4
such that the data approach optimally the limit
$S_\parallel(0)\rightarrow S(S+1)/3$ as $T\rightarrow\infty$.

Previous studies~\cite{birgeneau-77} of the 2D antiferromagnets
K$_2$NiF$_4$ ($S=1$) and K$_2$MnF$_4$ ($S=5/2$) show that within
$\sim$20\% of $T_N$ the critical magnetic scattering follows the
behavior predicted for the 2D Ising model, for which the exact
exponents are $\nu$=1 for the correlation length and $\gamma$=1.75 for
the susceptibility.  Our main purpose here is to study the 2D
Heisenberg regime, so the instrumental resolution was not optimized to
investigate the narrow, rapidly diverging peaks in the Ising critical
regime.  Even so, our fitted values for $\kappa_\parallel$ at
temperatures within 20\% of $T_N$ yield an exponent of $\nu=1.0\pm0.1$
in good agreement with the exact result $\nu=1$ for the 2D Ising
model.

%%==============================================================================
\begin{figure}
\centerline{\epsfxsize=3.1in\epsfbox{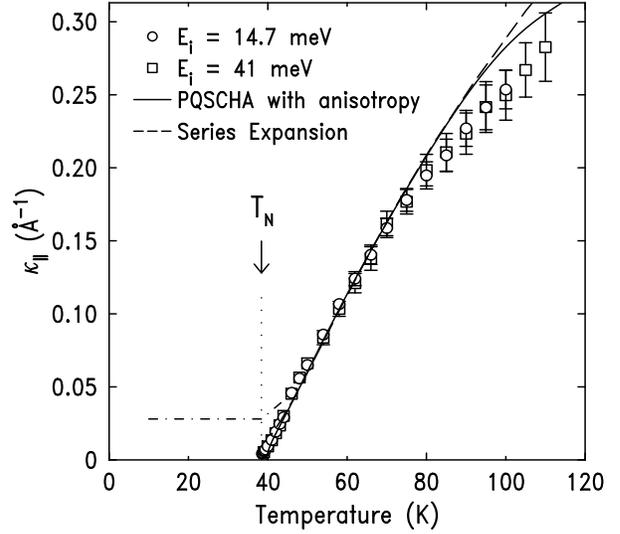}}
\vskip 5mm
\caption{Temperature dependence of $\kappa_\parallel$, the width
of the longitudinal spin component of the critical scattering.  At
high temperatures, $\kappa_\parallel$ exhibits 2D Heisenberg behavior,
crossing over to 2D Ising behavior near $T_N$.  The dash-dotted line
represents the interpolated temperature dependence of $\kappa_\perp$.
The solid line is the PQSCHA result [18] modified to include the
effects of the Ising anisotropy, and the dashed line is the prediction
from series expansion for the $S=5/2$ 2DSLQHA [7].  The plotted curves
have no adjustable parameters}
\label{Figure3}
\end{figure}
%%%==============================================================================

%%==============================================================================
\begin{figure}
\centerline{\epsfxsize=3.1in\epsfbox{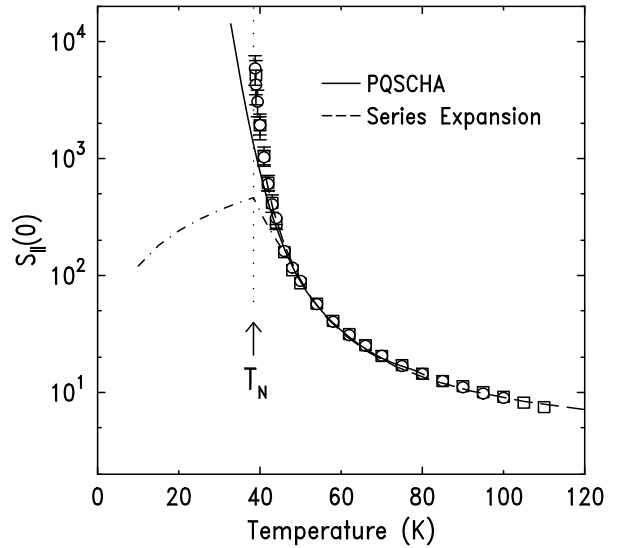}}
\vskip 5mm
\caption{Temperature dependence of $S_\parallel(0)$, the peak
intensity of the longitudinal critical scattering.  The dash-dotted
line represents the interpolated temperature dependence of
$S_\perp(0)$.  The solid line is the PQSCHA result [18], and the
dashed line is the prediction from series expansion for the $S=5/2$
2DSLQHA [7].  Since the neutron scattering intensity is not measured
in absolute units, the data are scaled to approach optimally the limit
$S_\parallel(0)\rightarrow S(S+1)/3$ as $T\rightarrow\infty$.  }
\label{Figure4}
\end{figure}
%%%==============================================================================

\section{Comparison with Theory}
\label{sec:4}

\begin{sloppypar}
A low-temperature theory for the 2DSLQHA was formulated by
Chakravarty, Halperin, and Nelson, in which they obtained the static
and dynamic properties of the 2DSLQHA by mapping it onto the 2D
quantum non-linear $\sigma$ model~\cite{chn}.  The 2D quantum
non-linear $\sigma$ model is the simplest continuum model which
reproduces the correct spin-wave spectrum and spin-wave interactions
of the 2DSLQHA at long wavelengths and low energies~\cite{chn,manou}.
First, CHN argued that the 2DSLQHA corresponds to the region of the 2D
quantum non-linear $\sigma$ model in which the ground state is
ordered -- the renormalized classical regime.  Then, CHN used
perturbative renormalization group arguments to derive an expression
for the correlation length to two-loop order, showing a leading
exponential divergence of $\xi$ versus inverse temperature.  Later,
Hasenfratz and Niedermayer employed chiral perturbation theory to
calculate the correlation length more precisely to three-loop
order~\cite{hn}.  In the renormalized classical scaling regime, the
correlation length is given by~\cite{hn}
\begin{equation}
\frac{\xi}{a}=\frac{e}{8}\frac{c/a}{2\pi\rho_s}~
e^{2\pi\rho_s/T}\left[1-\frac{1}{2}\left(\frac{T}{2\pi\rho_s}\right)+{\cal
O}\left(\frac{T}{2\pi\rho_s}\right)^2\right],
\end{equation}
which we refer to as the CHN-HN formula.  The parameters $\rho_s$ and
$c$ are the macroscopic spin-stiffness and spin-wave velocity of the
model, respectively.  For the nearest-neighbor 2DSLQHA, they are
related to the microscopic parameters $J_{nn}, S$ and the lattice
constant $a$ according to $\rho_s = Z_{\rho}(S) S^2 J_{nn}$ and $c =
Z_c(S)2\sqrt{2}aSJ_{nn}$.  The coefficients $Z_{\rho}(S)$ and $Z_c(S)$
are quantum renormalization factors, which can be calculated using
spin-wave theory ($S\geq1/2$), series expansion ($S=1/2, 1$), and Monte
Carlo techniques~($S=1/2$)~\cite{beard,hamer,wiese-ying}.  For
$S=1/2$, the spin-wave approximation \cite{hamer} gives
$Z_{\rho}(1/2)\simeq 0.699$ and $Z_c(1/2)\simeq 1.18$, whereas for
$S=5/2$, the factors $Z_{\rho}(5/2)\simeq 0.951$ and $Z_c(5/2)\simeq
1.03$ are closer to the classical limit of $Z_{\rho}=Z_c=1$.
\end{sloppypar}

\begin{sloppypar}
As mentioned above, the correlation length of the $S=1/2$ system
Sr$_2$CuO$_2$Cl$_2$ is well described by Eq.~(4) throughout the entire
experimental temperature range.  Subsequent Monte Carlo
work~\cite{beard} indicates that this agreement for the $S=1/2$ case
is, at least in part, coincidental; deviations exist, but they are too
subtle to be discerned experimentally.  In CHN-HN's formulation, the
natural expansion parameter for temperature is $T/2\pi\rho_s$.  For
all experimental systems, the lowest temperature at which 2D
Heisenberg behavior can be observed is bounded by a non-zero
temperature $T_N$ below which there is 3D long-range order.  For the
$S=1/2$ system Sr$_2$CuO$_2$Cl$_2$, the measured experimental
temperature range is $0.16<T/2\pi\rho_s<0.36$~\cite{greven}.  In our
present study of Rb$_2$MnF$_4$, which has a higher spin, though
smaller $J_{nn}$, the temperature range is similar,
$0.14<T/2\pi\rho_s<0.40$.  The Monte Carlo simulations of Beard
\et~\cite{beard} for $S=1/2$ indicate that the three-loop CHN-HN
formula, Eq.~(4), provides an accurate description of the $S=1/2$
2DSLQHA for temperatures
$T/2\pi\rho_s~{\Large\stackrel{_<}{_\sim}}~0.15$.  This barely
overlaps the temperature range studied experimentally here.  Further,
for $S=5/2$ the renormalized classical scaling regime is expected for
lower temperatures~\cite{elstner-series}.

In Fig.~5, we plot our results for the correlation length versus
inverse temperature.  Since $J_{nn}$ is known from independent
experiments~\cite{dewijn,cowley-spinwave}, there are no adjustable
parameters in the comparison between theory and experiment.  Also
plotted in Fig.~5 are the experimental data from Fulton {\it et
al.}~\cite{fulton} for the 2D $S=5/2$ Heisenberg antiferromagnet
KFeF$_4$.  The KFeF$_4$ system is less ideal because the nearest
neighbor spins in the planes are distorted from a square configuration
forming a rectangular lattice.  However, using the average of the
magnetic exchange coupling along the two different in-plane lattice
directions ($J_{nn}^{\rm avg}\simeq26.9$~K from neutron scattering
measurements of the spin-wave dispersion~\cite{fulton} with zero point
correction~\cite{hamer}), we find that the agreement for the measured
correlation lengths between Rb$_2$MnF$_4$ and KFeF$_4$ is excellent.
The reduced Ising anisotropy for KFeF$_4$ of $\alpha_I\simeq0.0045$ is
fortuitously almost identical to that for Rb$_2$MnF$_4$.  The
three-loop CHN-HN formula is plotted for $S=5/2$, and it appears as a
straight line on this logarithmic scale.  We used the spin-wave theory
values of $\rho_s \simeq 5.94 J$ and $c \simeq 7.30 Ja$~\cite{hamer}.
It is evident that the data deviate significantly from the three-loop
CHN-HN result for the quantum non-linear $\sigma$ model.
\end{sloppypar}

%%==============================================================================
\begin{figure}
\centerline{\epsfxsize=3.5in\epsfbox{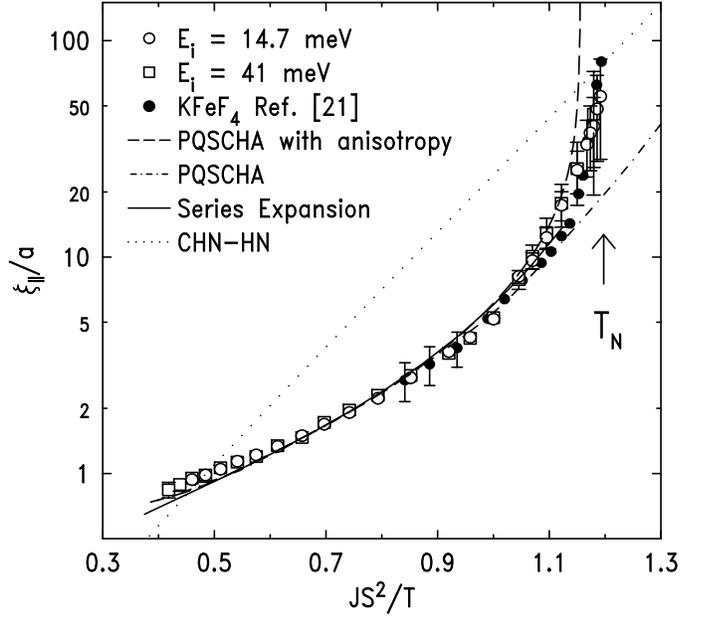}}
\vskip 5mm
\caption{The longitudinal correlation length $\xi_\parallel$ =
1/$\kappa_\parallel$ versus inverse temperature in reduced units.  The
open symbols are our data for Rb$_2$MnF$_4$.  The closed circles are
results for KFeF$_4$ from Ref.[21].  The dashed line is the PQSCHA
[18] results modified to include the Ising anisotropy.  The
dash-dotted line is the unmodified PQSCHA result.  The solid line is
the result of series expansion for the $S=5/2$ 2DSLQHA [7].  The
dotted line is the three-loop CHN-HN formula, Eq.~(4) [4,5].  The
comparisons are in absolute units, with no adjustable parameters}
\label{Figure5}
\end{figure}
%%==============================================================================

Figure~5 also includes the series expansion result for the $S=5/2$
2DSLQHA from Ref.~\cite{elstner-series}.  There is good agreement
between the series expansion prediction, with plotted correlation
lengths up to 14 lattice constants, and our data over most of the
temperature range.  Systematic deviations exist at very high
temperatures, where the correlation length is small, and at low
temperatures near $T_N$, where the correlation length begins to
diverge rapidly.  At high temperatures, where the correlation length
decreases to one lattice constant or less, $\xi$ is not precisely
defined.  Also, at such high temperatures where the measured peaks
become broad in reciprocal space, the use of an Ornstein-Zernike
Lorentzian lineshape may be called into question, and the background
is more difficult to define.  Accordingly, it is not surprising that
slight deviations exist at high temperatures.  We have already pointed
out that Ising fluctuations will dominate near $T_N$, so the
correlation length will necessarily deviate from Heisenberg behavior
in the immediate vicinity of $T_N$.

An alternative theoretical analysis of the 2DSLQHA has been carried
out by Cuccoli \et~\cite{cuccoli} in which they treat quantum
fluctuations in a self-consistent Gaussian approximation separately
from the classical contribution.  In their approach, which they label
the purely-quantum self-consistent harmonic approximation (PQSCHA),
the quantum spin Hamiltonian is rewritten as an effective classical
Hamiltonian, where the temperature scale is renormalized due to
quantum fluctuations, and the new classical spin length appears as
$S+\frac{1}{2}$.  Defining the reduced temperature as
$t=T/\{J_{nn}(S+\frac{1}{2})^2\}$, the correlation length for the
2DSLQHA is then simply given by
\begin{equation}
\xi(t) = \xi_{cl}(t_{cl})~~~{\rm with}~~~t_{cl} =
\frac{t}{\theta^4(t)}.
\end{equation}
Here, $\xi_{cl}$ is the correlation length for the corresponding
classical 2D square-lattice Heisenberg model, and $\theta^4(t)$ is a
temperature renormalization parameter.  The PQSCHA is most accurate in
the limit where the quantum fluctuations are weak, and correspondingly
$\theta^4(t)$ is near unity.  The calculations of Cuccoli
\et~\cite{cuccoli} show that this is the case for $S=5/2$ over an
extended temperature range; further, their results show good agreement
with the existing experimental $S=1/2$ and $S=1$ data over the
appropriate high-temperature ranges.  The behavior of $\xi_{cl}$ is
determined from classical Monte Carlo simulations.  The PQSCHA result
for $S=5/2$ is plotted in Fig.~5 as the dash-dotted line.  Similar to
the series expansion result, there is good agreement between the
PQSCHA theory and our data and those in KFeF$_4$ with no adjustable
parameters.  Again, as $T_N$ is approached, the data deviate from the
theoretical curve because of the crossover to Ising behavior.

\begin{sloppypar}
Keimer \et~\cite{keimer-214} have derived a simple mean-field model
which allows one to incorporate the effects on the longitudinal spin
correlations of the staggered Ising anisotropy field as appears in
Eq.~(1).  In this model, the unperturbed Heisenberg correlation length
is replaced by the form
\begin{equation}
\xi(\alpha_I,
T)=\frac{\xi_{Heis}(T)}{\sqrt{1-\alpha_I\xi^2_{Heis}(T)}}.
\end{equation}
Using the value $\alpha_I$=0.0047 measured in Rb$_2$MnF$_4$ and the
form for $\xi_{Heis}(T)$ given by the PQSCHA for the Heisenberg model,
we obtain the dashed line shown in Fig.~5.  This mean-field result for
the correlation length nicely captures the crossover from Heisenberg
to Ising spin correlations in both Rb$_2$MnF$_4$ and KFeF$_4$.  By
design, this model takes into account the growing importance of the
local anisotropy field as the spin correlation length grows; it does
not include 2D Ising critical effects.  Thus, the predicted divergence
is mean-field-like with a power-law exponent of $\nu$=1/2 instead of
the 2D Ising result of $\nu$=1.  This rapid mean-field divergence is
also elegantly seen as the ``PQSCHA-with-anisotropy'' curve rises
above our data very close to $T_N$.  Nevertheless, this mean-field
model with no adjustable parameters predicts $T_N$ to within 3\% in
Rb$_2$MnF$_4$ and to within 6\% in KFeF$_4$.  The accuracy of this
model is highlighted when compared to the conventional mean-field
theory prediction of $T_N=4J_{nn}S(S+1)/3$ which is higher than the
experimentally measured transition temperatures by more than a factor
of two.
\end{sloppypar}

\begin{sloppypar}
In Fig.~6 we plot the ratio $S_\parallel(0)/\xi_\parallel^2$ versus
temperature for Rb$_2$MnF$_4$.  The open symbols indicate data for
which the Ising critical fluctuations become significant, so we
concentrate here on the closed symbols in the temperature regime where
the behavior is predominantly Heisenberg-like.  The low-temperature
analysis of CHN-HN predicts that this quantity should follow the form
\begin{equation}
\frac{S(0)}{\xi^2}=A2\pi M_s^2\left(\frac{T}{2\pi\rho_s}\right)^2
\left[1+C\frac{T}{2\pi\rho_s}+{\cal O}\left(\frac{T}{2\pi\rho_s}\right)^2\right],
\end{equation}
where $M_s$ is the $T=0$ staggered magnetization and $A$ and $C$ are
universal constants.  The same form, which is written above at
three-loop order, can also be derived from a renormalization group
analysis of the classical model~\cite{elstner-long,nelson}.  In an
alternative approach, Kopietz~\cite{kopietz} obtained the leading
$T^2$ behavior based on a Schwinger boson mean-field theory~\cite{aa}.
Using the values for the universal numbers $A$ and $C$ in Eq.(7)
obtained from Monte Carlo calculations on the $S=1/2$ and $S=1$
models ($A=4.5$ and $C=0.5$)~\cite{kim,harada}, we plot Eq.(7) in
Fig.~6.  It appears that the low-temperature data points in the
Heisenberg regime may overlap with the $T^2$ law; however, the data
depart significantly from the predicted $T^2$ law at higher
temperature.  Since the overall scaling factor for the
$S_\parallel(0)$ data was determined in Fig.~4, there are no
adjustable parameters in this plot.  Also plotted are results from
series expansion~\cite{elstner-series} and the PQSCHA
theory~\cite{cuccoli} for the $S=5/2$ 2DSLQHA.  There is a rough
overall agreement with both of these latter results, but systematic
deviations clearly exist.
\end{sloppypar}

%%==============================================================================
\begin{figure}
\centerline{\epsfxsize=3.1in\epsfbox{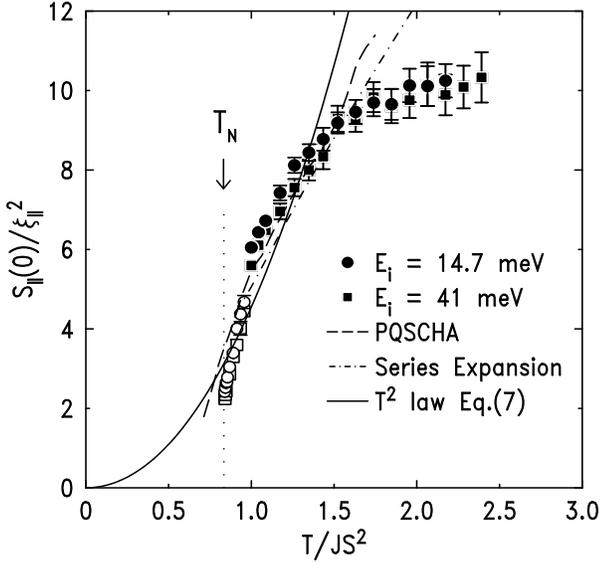}}
\vskip 5mm
\caption{The ratio $S_\parallel(0)/\xi_\parallel^2$ versus
temperature.  The dashed line is the PQSCHA result [18].  The
dash-dotted line is the prediction from series expansion for the $S=5/2$
2DSLQHA [7].  The solid line is Eq.(7) with the parameters $A$ and
$C$ obtained from [26].  }
\label{Figure6}
\end{figure}
%%==============================================================================

\section{Summary}
\label{sec:5}

In summary, we find that the instantaneous spin-spin correlations in
Rb$_2$MnF$_4$ are quantitatively described by high-temperature series
expansion results and the PQSCHA theory for the $S=5/2$ 2DSLQHA with
no adjustable parameters.  There is also good absolute agreement with
previous results for KFeF$_4$, which is a somewhat less ideal $S=5/2$
2DSLQHA.  Our data, which correspond to correlation lengths up to
about 50 lattice constants, are not well described by the three-loop
CHN-HN prediction based on the 2D quantum non-linear $\sigma$ model.
This discrepancy almost certainly derives from the fact that the
experimental data correspond to temperatures where the $S=5/2$ 2DSLQHA
is not in the renormalized classical low-temperature regime described
by Eq.~(4), but rather in the classical scaling regime.  Arguments by
Elstner \et~\cite{elstner-series}, based on CHN's use of cutoff
wavevectors for integrations over the Brillioun zone, give a crossover
temperature $T_{cr} \sim J_{nn}S$ between renormalized classical
behavior at low-temperature and classical scaling behavior at
high-temperature.  For Rb$_2$MnF$_4$, $T_{cr} \sim 18.4~$K, which
suggests that the experimental data lie in the classical scaling
regime.  In addition, the PQSCHA theory predicts the correlation
length accurately in absolute units for this material, which is
consistent with the statement that quantum fluctuations are not large
and the system is in the classical scaling regime in the temperature
range measured.

As of yet, no quantum Monte Carlo results exist for the $S=5/2$
2DSLQHA.  In Monte Carlo work for $S=1/2$~\cite{beard,kim} it was
found that $\xi$ deviates somewhat from the three-loop CHN-HN result
for correlation lengths less than $\sim1000$ lattice constants.
However, results for the ratio $S(0)/\xi^2$~\cite{kim} with fitted
universal constants show quantitative agreement with the $T^2$ law of
Eq.~(7) over the complete temperature range.  Recent Monte Carlo work
for $S=1$~\cite{harada} shows that for correlation lengths of less
than 25 lattice constants, $\xi$ is not well described by the
three-loop CHN-HN low-temperature formula.  However, the $S=1$ Monte
Carlo data match well both the results from the PQSCHA theory for
$S=1$ and the experimental data for La$_2$NiO$_4$.  Also, the leading
$T^2$ behavior for $S(0)/\xi^2$ is followed by the Monte Carlo data for
$S=1$.  This is in contrast with our results for Rb$_2$MnF$_4$ where
we find a significant departure from the leading $T^2$ behavior for
$S(0)/\xi^2$ at high temperatures or, equivalently, short correlation
lengths.

\begin{sloppypar}
Finally, we note that the 2D Heisenberg-Ising crossover behavior in
both Rb$_2$MnF$_4$ and KFeF$_4$ are predicted reasonably well by
Eq.~(6) without any adjustable parameters.  However, since Eq.~(6) is
a mean-field result it cannot account for the asymptotic 2D Ising
critical behavior.
\end{sloppypar}

In conclusion, we now have a quite complete understanding of the spin
correlation length in the $S=5/2$ 2D Heisenberg antiferromagnet for
the experimentally relevant temperature range.  Further work is
required to understand the static structure factor peak intensity
$S(0)$ to the same degree.

\

We thank P. Verrucchi and R.R.P. Singh for sending us the numerical
PQSCHA and series expansion results, respectively.  We thank U.-J.
Wiese for helpful discussions.  The work at MIT was supported by the
National Science Foundation - Low Temperature Physics Program under
award number DMR 97-04532.  Work at Brookhaven National Laboratory was
carried out under Contract No. DE-AC02-98CH10886, Division of
Materials Science, U.S. Department of Energy.

%%%%%%%%%%%%%%%%%%%%%%%%%%%%%%%%%%%%%%%%%%%%%%%%%%%%%%%%%%%%%%%%%%%%

\end{document}